\documentclass[showpacs,nofootinbib]{revtex4}
\usepackage{amssymb}

\usepackage{natbib}
\usepackage{graphicx}

\def\allsum{\mathrm{{\sf{}S}}'}

\begin{document}
\title{An empirically constructed dynamic electric dipole polarizability function of magnesium and its applications}
\author{James F. Babb}
\affiliation {Harvard-Smithsonian Center for Astrophysics, 60 Garden St., MS 14, Cambridge, MA 02138}
\date{\today}
\begin{abstract}
The dynamic electric dipole polarizability function for the magnesium atom is
formed by assembling the atomic electric dipole oscillator strength distribution from
combinations of theoretical and experimental data for resonance oscillator strengths 
and for photoionization cross sections of valence and inner shell
electrons.
Consistency 
with the oscillator strength (Thomas-Reiche-Kuhn) sum rule
requires the adopted principal resonance line oscillator strength 
to be several percent lower 
than the values given in two critical tabulations, though the value adopted is 
consistent with a number of theoretical determinations.
The static polarizability is evaluated.
Comparing the resulting dynamic polarizability as a function of photon energy with
more elaborate calculations reveals
the contributions of inner shell electron excitations.
The present results are applied to calculate 
the long-range interactions between two and three magnesium atoms and the interaction 
between a magnesium atom and a perfectly conducting metallic plate.
Extensive comparisons of prior results for the principal resonance line oscillator strength, for the static polarizability,
and for the van der Waals coefficient are given in an Appendix.
\end{abstract}

\maketitle
%
%
\section{Introduction}
Magnesium is an abundant element currently of interest in several applications.
Analysis of photo association spectroscopy for the Mg dimer~\cite{TieKotJul02,KnoRuhTie13} indicates that the $s$-wave scattering length 
for collisions between two ground state Mg atoms is 
positive~\cite{TieKotJul02},  with the accuracy of the determination affected by 
the remnant uncertainty in
the value of the atom-atom van der Waals constant~\cite{KnoRuhTie13}. 
(A Bose-Einstein condensate of Mg atoms has not been created experimentally, to date.)
However, a determination of the 
leading term in the long-range interaction of electronically excited states of the Mg dimer,
which is related to the principal resonance line oscillator strength, using molecular spectroscopy has
been elusive~\cite{KnoRuhTie14}.
Mg atoms were investigated as a possible sympathetic cooling agent in collisions with NH~\cite{SolZucHut09,GonMarMay11}
and Lonij~\textit{et al.}~\cite{LonKlaHol11} theoretically explored the interaction of an Mg atom with a wall for
applications to atom interferometry.
And, while Mg is abundant in the solar system,
it is interesting to note that the Mg principal resonance line was recently detected in the observation of the exoplanet HD~209458b 
using transit spectroscopy ~\cite{VidMadHui13}.
Modeling  Mg absorption in exoplanet atmospheres depends proportionally on the principal resonance line absorption oscillator
strength~\cite{BouLecVid14}, the value of which in turn affects the use of 
that line as a probe of escaping atoms in exoplanet atmospheric spectroscopy~\cite{BouLecVid15}.

An analysis of the dynamic electric dipole polarizability function of Mg is valuable for several reasons.
First, because the principal resonance line oscillator strength is an important contributor to the function,
it is possible to determine a value that achieves consistency with oscillator strength sum rules.
Secondly, it is desirable to have an independent assessment of the completeness of existing elaborate calculations of
the function itself, for which extensive tabulations from two different types of calculations are available~\cite{DerPorBab10,JiaMitChe15}.

Calculations of dynamic electric dipole polarizabilities are of intrinsic theoretical
interest due to the challenges  inherent in treating 
correlations and excitations of all electrons quantum-mechanically at different photon energies~\cite{Amu90,AmuCheYar12}.
Such calculations are necessarily important benchmarks for theoretical methods
applied to  photoabsorption~\cite{Ber02}, photodetachment~\cite{MasSta00},
blackbody radiation shifts~\cite{MidFalLis12,SheLemHin12,BelSheLem12} and AC Stark shifts~\cite{StaBudFre06}, 
magic wavelengths~\cite{BarStaLem08}, and
parity non conservation amplitudes~\cite{Khr91}, as well as being helpful 
in the ongoing development of density functional theory (DFT) methods for dispersion forces 
(cf.~\cite{BasHesSal08,TkaSch09,TkaDiSCar12,TaoPerRuz13,TouRebGou13}). 
In addition, for metals experimental data at a wide spectrum of photon energies are relatively scarce
though X-ray  data, and sometimes optical data~\cite{GoeHoh95,SarBeiShe06},  are available.
There is recent progress for systematically measuring static polarizabilities~\cite{MaIndZha15}.
Many theoretical approaches are available, but their reliability in calculating dynamic polarizabilities
can be difficult to gauge without critical evaluation, but
critical evaluations are limited to the static polarizabilities~\cite{Sch06,ThaLup06,MitSafCla10}.
Nevertheless, dynamic polarizabilities are of great utility in calculating 
coefficients appearing in certain potential energies, particularly van der Waals constants (for investigations of
ultra-cold collisions, for photo association spectroscopy, 
and for ultra-cold gas studies) and Lennard-Jones constants (for 
atom-surface interactions, where recent
applications include 
tests for gravity-related new physics at submillimeter distances~\cite{DimGer03,HarObrMcG05,MurHau09},
optical clocks~\cite{DerObrDzu09,DzuDer10}, atom-graphene interactions~\cite{AroKauSah14}, noncontact van der Waals friction~\cite{JenLacDeK15},
and interactions between nanostructures~\cite{TaoPer14}).

In a previous paper treating the sodium atom~\cite{KhaBabDal97}, 
a semi-empirical theory utilizing oscillator strength sum rules  
and input data from experiments and  calculations
predicted a value of  the van der Waals coefficient~\cite{KhaBabDal97}, which was found to be in harmony 
with subsequent experimentally determined fits from photo-association spectroscopy data~\cite{vanVer99,KnoSchSce11}
and  \textit{ab initio} theoretical methods~\cite{DerJohSaf99}.
In the case of sodium, the availability of precise measurements
of the principal resonance line oscillator strength from photo-association
spectroscopy and of the static electric dipole polarizability from atom interferometry 
augmented the semi-empirical analysis~\cite{KhaBabDal97}.
In the case of magnesium, such data are not  available.
Therefore, in this paper, for Mg, the electric dipole oscillator strength distribution is composed 
using extant data on electric dipole oscillator strengths, photoabsorption cross sections, and energies
obtained experimentally and theoretically. 
As will be shown, consistency with the oscillator strength (Thomas-Reiche-Kuhn) sum rule requires a value of the principal
resonance line oscillator strength that is several percent lower than values
listed in critical tabulations by Morton~\cite{Mor03} and by Kelleher and Podobedova~\cite{KelPod08}.
Other evidence for the value adopted is given.
The static electric dipole polarizability is evaluated and compared with other values.
The  dynamic polarizability function is calculated and compared with previous
results obtained by Porsev et al.~\cite{DerPorBab10} using 
configuration interaction and 
many-body perturbation theory with core contributions (CI-MBPT)~\cite{PorDer06}
and by Jiang \textit{at al.}~\cite{JiaMitChe15} using the configuration 
interaction with semi-empirical core-valence interaction (CICP) method.
The present dynamic polarizability function  is used to evaluate the  van der Waals constant,  Axilrod-Teller-Muto constant, 
and atom-surface interactions.
Results from the literature for the principal resonance line oscillator strength,
static electric dipole polarizability, and van der Waals constant are collected and compared in the Appendix.

%
%
\section{Dipole oscillator strength sum rules}

The absorption oscillator strength from the ground state $|0\rangle$ with eigenvalue $E_0$ to
an excited state $|n\rangle$ with eigenvalue $E_n$ is 
\begin{equation}
f_n = {\textstyle{\frac{2}{3}}}(E_n-E_0) \left|\left\langle 0 \left| \sum_{i=1}^N  {\mathbf r}_i \right| n\right\rangle\right|^2,
\end{equation}
where ${\mathbf r}_i$ is the position vector of electron $i$, and $N$ is the number of electrons.
Atomic units are used throughout unless otherwise specified.

Denoting by $\allsum_n$ the sum-integral (the sum over all discrete transitions excluding the initial state and the integration over 
all continuum states), the resultant sum rules are
\begin{equation}
S(0) = \allsum_n f_n = N ,
\end{equation}
with $N=12$ for Mg,
\begin{equation}
S(-1) = \allsum_n f_n / (E_n-E_0) =
  {\textstyle{\frac{2}{3}}} \left\langle 0 \left| \left(\sum_{i=1}^N  {\mathbf r}_i \right)^2\right| 0 \right\rangle ,
  \end{equation}
and 
\begin{equation}
S(-2) =\allsum_n f_n / (E_n-E_0)^2 = \alpha(0) ,
\end{equation}
where
$\alpha(0)$ is the static electric dipole polarizability.
The 
dynamic electric dipole polarizability function is 
\begin{equation}
\alpha(\omega) = \allsum_n \frac{f_n}{(E_n-E_0)^2 -\omega^2 },
\end{equation}
where $\omega$ is the photon energy. 
By direct integration 
the $S(-1)$ sum rule is related to the atom-wall interaction coefficient $C_3$~\cite{KhaBabDal97}
\begin{equation}
\label{atom-wall-formula}
C_3 = {\textstyle{\frac{1}{8}}} S(-1) = \frac{1}{4\pi}\int_0^\infty d\omega \,\alpha(i \omega),
\end{equation}
the van der Waals coefficient is
\begin{equation}
\label{vdW-formula}
C_6 = \frac{3}{\pi}\int_0^\infty d\omega\,[\alpha (i \omega)]^2 ,
\end{equation}
and the Axilrod-Teller-Muto coefficient is
\begin{equation}
\label{Axilrod-Teller-formula}
C_9 = \frac{3}{\pi} \int_0^\infty d\omega\;[\alpha (i\omega)]^3.
\end{equation}

\section{Oscillator strength distribution}\label{OSD}

A magnesium atom has twelve electrons.
Their  configuration is $(1s^2 2s^2 2p^6 3s^2 )\;{}^1S_0$.

\subsection{Discrete transitions}\label{discrete}
Sources with tabulations of values, experimental and theoretical, for the absorption
oscillator strengths are Mitchell~\cite{Mit75a}, Mendoza and Zeippen~\cite{MenZei87a},
Ray and Mukherjee~\cite{RayMuk89},
J\"onsson and Fischer~\cite{JonFis97}, Hamonou and Hibbert~\cite{HamHib08}, and Derevianko and Porsev~\cite{DerPor11}.

There are numerous theoretical determinations of the oscillator strength for the 
principal resonance line $(3s^2)\; {}^1S - (3s3p) \;{}^1P^o$. 
A detailed survey is given in the Appendix.
Reliable theoretical calculations range from 1.709 to 1.76,
and there were at least ten experimental determinations as of 2003~\cite{Mor03}.
In a critical review, Morton~\cite{Mor03} adopted a value of $1.83\pm 0.03$ based on a weighted mean
of the ten experimental values.
A long-standing discrepancy between experimental and theoretical trends was noted previously~\cite{Fis75,VicSteLau76,JonFisGod99,ZatBarGed09}.
The tendency of
high-level theoretical results to be less than 1.8  was noted recently by Zatsariny \textit{et al.}~\cite{ZatBarGed09} who calculated a nonrelativistic
value of 1.738 
and  pointed out that a ``very extensive
and essentially converged multiconfiguration Hartree-Fock (MCHF)'' 
\textit{ab initio} calculation by J\"onsson, Fischer, and Godefroid~\cite{JonFisGod99}  found 1.717.
J\"onsson, Fishcher, and Godefroid~\cite{JonFisGod99},  using the observed transition energy
to evaluate the oscillator strength, obtained 1.710 and also pointed out (see their Table 11) that (as of  
1999) theoretical values were consistently smaller than the experimental ones (to date) by about 5 percent.
Recently, Derevianko and Porsev~\cite{DerPor11} quote for the matrix element governing the line strength
a value $4.03 \pm 0.02$ with error of 0.5\% based on their
calculations from 2001~\cite{PorKozRak01}; the corresponding oscillator strength
using the experimental transition energy~\cite{KelPod08} is $1.73 \pm 0.02$.
The original NIST (NBS) tabulation of 1969~\cite{WieSmiMil69} adopts $1.8\pm 0.18$ from an average of 
the experiments of Refs.~\cite{Lur64,SmiGal66} and the calculation of Weiss~\cite{Wei67}. The
value 1.8 (cited as a private communication from A.~W. Weiss) is given with 3 percent error $(\pm 0.05)$
at the 90 percent confidence level in the 2008 NIST revised tabulation~\cite{KelPod08}.

I adopt the value 1.75, which is 
the lower limit of the value  1.80(5) from 
Kelleher and Podobedova~\cite{KelPod08} and the upper limit
of the value $1.73(2)$ recommended by Derevianko and Porsev.

Mitchell~\cite{Mit75a} used the anomalous dispersion (hook) method to measure the second $(3s$--$4p)$ through sixth $(3s$--$8p)$ resonance
transition oscillator strengths and found, respectively, 
$0.107\pm 0.0019$, 
$(2.27\pm 0.12)\times10^{-2}$,
$(8.53\pm 0.46)\times10^{-3}$,
$(4.11\pm 0.36)\times10^{-3}$, and 
$(2.34\pm 0.15)\times10^{-3}$,
all determined relative to the value of 1.72 for the principal transition that
Victor and Laughlin~\cite{VicLau73} calculated
using a  semi-empirical model potential method.
Other theoretical determinations of $f$ values for the second and higher resonance transitions are those of 
Saraph~\cite{Sar76}, Mendoza and Zeippen~\cite{MenZei87a}, Chang and Tang~\cite{ChaTan90}
and Zatsarinny \textit{et al.}~\cite{ZatBarGed09}.
I adopt the values of Chang and Tang, who calculated
for the principal to sixth resonance lines, respectively, oscillator strengths 1.75, 0.111, 0.024, 0.0091, 0.0043, and 0.0024.
These five discrete (second to sixth resonance) transitions contribute $0.151$ to the $S(0)$ sum 
and $2.81$ to the $S(-2)$ sum for the $3s$ shell. 

Including the principal resonance line, the  \textit{discrete transition} contribution from the $3s$ shell to $S(0)$  is 1.90 
and to $S(-2)$ is $71.4$.

\subsection{Continuum transitions}

A number of sources exist for the continuum oscillator strengths
corresponding to the ejection of a $3s$ electron~\cite{DitMar53,BurSea60,BatAlt73,ParReeTom76,DesMan83,PreBurGar84,RadJoh85,YehLin85,FisSah87,MenZei87b,MocSpi88b,Alt89,VerYakBan93,ChiHua94,FunYih01,KimTay00,WehLukJur07,HauKamKos88,WanWanZho10,PinBalAbd13,LeeBalAbd15}.
The photoionization cross sections calculated using a variational MCHF method by Fischer and Saha~\cite{FisSah87}
are in good agreement with the experimental results of Wehlitz \textit{et al.}~\cite{WehLukJur07}. 
The threshold cross section of 2.5~Mb calculated by  Fischer and Saha~\cite{FisSah87}
is slightly larger than both the value $2.1\pm0.3$~Mb measured by Fung \textit{et al.}~\cite{FunYih01} 
and the value $2.36\pm 0.02$~Mb that Parkinson,
Reeves, and Tomkins~\cite{ParReeTom76} found
by extrapolation from the measured discrete oscillator strengths of Ref.~\cite{Mit75a}.
Wehlitz \textit{et al.}~\cite{WehLukJur07}  normalized  their own measurements to the value 2.1~Mb at threshold.
At a photon energy of 30~eV, a recent calculation by Pindzola \textit{et al.}~\cite{PinBalAbd13} using 
a time-dependent close-coupling method with an effective core potential 
gives $\sigma_{3s} (30\,\mathrm{eV})$ $=$ $0.217\;\mathrm{Mb}$, while
Verner~\textit{et al.}~\cite{VerYakBan93} calculate $0.255\;\mathrm{Mb}$.
At 80~eV, the ``complete''  experiment of Haussman \textit{et al.}~\cite{HauKamKos88},
for which the total absorption cross section of Ref.~\cite{HenLeeTan82} was used for normalization, yields 
$\sigma_{3s} (80\,\mathrm{eV})$ $=$ $0.080\pm 0.011\;\mathrm{Mb}$
from the main transition compared to the value  $0.087$ from Verner~\textit{et al.}
Haussman \textit{et al.} also measured an additional $0.014 \pm 0.004\;\mathrm{Mb}$
contribution from satellites.
The measurements of  Wehlitz \textit{et al.}~\cite{WehLukJur07} are adopted for energies from threshold to 11.6~eV
and  the $3s$ cross section data were extended to higher energies using the results of Verner~\textit{et al.}~\cite{VerYakBan93}.

The contributions to the three sum rules $S(0)$, $S(-1)$, and $S(-2)$ 
from the continuum 
are, respectively, 0.261, 0.162, and 0.277. 
Combining these with the discrete contributions from the Sec.~\ref{discrete} the total
valence shell contributions are 2.16, 11.8, and 71.7,
which are listed in Table~\ref{osc}.
The calculated valence shell contribution of 2.16 to the $S(0)$ sum confirms that the contribution from the $3s$ electron 
to $S(0) $ is greater than 2~\cite{Stw71,ResCurBro08},  indicating configuration
interaction of the valence electrons with the core electrons.  Maeder and Kutzelnig~\cite{MaeKut79}
obtained 2.06 using a model potential including core-valence correlation.
 
The excitation and ejection of $K$ shell electrons was considered by Verner~\textit{et al.}~\cite{VerYakBan93},
Kutzner \textit{et al.}~\cite{KutMayTho02}, and Haso\u{g}lu \textit{et al.}~\cite{HasAbdNab14}.
Haso\u{g}lu \textit{et al.} used  $R$-matrix methods to calculate excitation of $K$ shell electrons 
to $np$ states resonant below the threshold. The resonances
are estimated to contribute 0.03 to the value of $S(0)$.
The  relativistic random-phase approximation
modified to include relaxation effects (RRPAR) calculations of Kutzner \textit{et al.} are 
in good agreement with the  cross sections
calculated by Verner~\textit{et al.} from threshold to high energies.
Banna~\textit{et al.}~\cite{BanSlaMat82} measured
the shake-up peak just above the 
threshold, but absolute cross sections are not available.
The cross sections of Verner are adopted from 
the threshold for ejection of a $1s$ electron at 1310.9~eV.
The $K$ shell contribution yields 1.56 to $S(0)$ and it is negligible for $S(-1)$ and $S(-2)$.
The $1s$ contribution to $S(0)$ found here for Mg is comparable to that found for Na~\cite{KhaBabDal97}.

Deshmukh and
Manson~\cite{DesMan83}, Nasreen, Manson, and Deshmukh~\cite{NasManDes89},  Kutzner \textit{et al.}~\cite{KutMayTho02}, and  Verner \textit{et al.} \cite{VerYakBan93} calculated 
the partial cross sections for the ejection of a $2s$ electron.
The cross section value 0.3~Mb at the threshold energy of 94.0~eV
is adopted~\cite{NasManDes89,KutMayTho02} and linearly joined to the results of Verner \textit{et al.} at 270~eV,
which are used for higher energies.
For the $2s$ shell the contribution to the sums $S(0)$, $S(-1)$,
and $S(-2)$ are, respectively,  1.02, 0.1, and 0.02.

The remaining oscillator strength must come from the $2p$ shell
and the expected contribution to $S(0)$ is 7.26.

Deshmukh and
Manson~\cite{DesMan83}, Altun~\cite{Alt89}, Nasreen, Manson, and Deshmukh~\cite{NasManDes89},
Kutzner, Maycock, and
Thorarinson~\cite{KutMayTho02}, 
investigated photoionization of a $2p$ electron.
There a number of resonances corresponding to excitation to autoionizing states~\cite{Alt89}
and they contribute significantly to the oscillator strength.
Measurements of the cross section by Haussman \textit{et al.}~\cite{HauKamKos88} at 80~eV
found that the resonances constitute  25\% of the $2p$ shell photoionization cross section.
The cross sections averaged over the resonances from the
correlated length gauge  many-body 
perturbation theory (MBPT) calculations of Altun (Figure 9 of Ref.~\cite{Alt89}) are adopted from  63.29 eV to
344.89~eV, giving  
 $\sigma_{2p} (80\,\mathrm{eV})$ $=$ $6.2\;\mathrm{Mb}$ which is slightly larger than 
the reference value used by Haussman \textit{et al.}
The calculations of Verner \textit{et al.} are used from the threshold energy of 54.9~eV up to 63.29~eV
and for energies above 344.89~eV.
These data yield 
for the $2p$ shell the contributions to the sums $S(0)$, $S(-1)$,
and $S(-2)$, respectively,  7.33. 1.61, and 0.45.
If the calculations of Altun are multiplied by a factor of 0.985, the cross section
at 80~eV becomes $6.1$~Mb and the sums
$S(0)$, $S(-1)$, and $S(-2)$ are calculated to be, respectively,
7.26, 1.60, and 0.44 and the values are listed in Table~\ref{osc}.

The contributions to the sums $S(0)$, $S(-1)$, and $S(-2)$ from the excitation of the $1s$, $2s$, $2p$,
and $3s$ electrons are summarized in Table~\ref{osc}.

\begin{table*}
\caption{\label{osc} Contributions to the sum rule $S(k)$ for  Mg. 
}
\begin{ruledtabular}
\begin{tabular}{llll}
\multicolumn{1}{l}{ }  & \multicolumn{1}{l}{$S(0)$} &
\multicolumn{1}{l}{$S(-1)$} & \multicolumn{1}{l}{$S(-2)$} \\
\hline
1s		& 1.56 	& 0.02	& $...$    	\\
2s		& 1.02 	& 0.10	& 0.02	\\
2p		& 7.26 	& 1.60	& 0.44	\\
3s		& 2.16	& 11.8	& 71.7      \\
Total	& 12.0	& 13.5	& 72.2	\\
\end{tabular}
\end{ruledtabular}
\end{table*}

\subsection{Discussion}

Stwalley~\cite{Stw71}, Pal'chikov and Ovsiannikov~\cite{PalOvs04}, 
Ovsiannikov~\textit{et al.}~\cite{OvsPalKat06}, and Sarkisov~\textit{et al.} \cite{SarBeiShe06}
constructed  oscillator strength distributions of Mg for calculations of dynamic polarizabilities,
considering valence transitions. Sarkisov~\textit{et al.} \cite{SarBeiShe06} included an
estimate of $2p$ excitations.

From their tabulated data, the results of Pal'chikov and Ovsiannikov~\cite{PalOvs04}  and 
Ovsiannikov~\textit{et al.}~\cite{OvsPalKat06} indicate  
a total discrete contribution of $1.9$ to the $S(0)$ sum rule for $3s$ discrete transitions
in agreement with the present result.
They used a value of $1.73$ for the principal resonance line oscillator strength, which 
offsets in the sum rule their slightly larger value of $0.122$ for the second resonance line oscillator strength,  
compared to the present adopted values of, respectively, $1.75$ and $0.111$.
For the $S(-2)$ sum, they find 
for the second to sixth resonance transitions a contribution of $3.12$ to $S(-2)$, to be compared to the present value of $2.81$.
The difference between their value and the present value is due primarily to the different values of the 
second resonance line oscillator strength.
Their total value (valence electrons) for $S(-2)$ is 71.39, while the present value for the $3s$ shell is 71.7.

Including all shells, the present value of $\alpha(0)$ is 72.2 from the $S(-2)$ sum rule, see Table~\ref{osc}. 
It lies only 0.2 above  
the range of values $71.3(7)$ recommended by Porsev and Derevianko~\cite{PorDer06}
and it is compared with a number of other theoretical calculations  in the Appendix.

Note that if the value 1.8 is adopted for the principal oscillator strength~\cite{KelPod08}, without
any other adjustments to the adopted data, the present $S(0)$ sum becomes 12.05
and $\alpha(0)$ becomes  $73.4$,
which is far beyond the value recommended by Porsev and Derevianko.
The value 1.83 for the oscillator strength adopted by Morton~\cite{Mor03} is more difficult to reconcile
within the present analysis.
The $S(0)$ sum becomes 12.08 and the  value of $\alpha(0)$  becomes 74.6.
Sarkisov~\textit{et al.}~\cite{SarBeiShe06} use the oscillator strength data from
Morton~\cite{Mor03} and estimate the $2p$ and $3s$ continuum contributions
using the data from Verner~\textit{et al.}~\cite{VerYakBan93} and  find $\alpha(0)=73.6$. 
Stwalley's early calculation~\cite{Stw71} used the value $1.82(5)$ for the principal resonance line oscillator strength
(see his Ref.~7 for sources) and obtained an estimate $\alpha(0)=75.0 \pm 3.0$.

A recent experiment~\cite{MaIndZha15} using a pulsed cryogenic
molecular beam  electric deflection method obtained a value $\alpha(0) = 59(15)$,
which is not sufficiently accurate to discriminate between theoretical calculations.

Reshetnikov~\textit{et al.}~\cite{ResCurBro08} explored the relationship between the uncertainty
in $\alpha(0)$ and the uncertainty in the lifetime of the first resonance transition in two-valence electron atoms and ions.
Their formalism allows a valence shell contribution to $N$ that is not exactly 2, as found here and in Ref.~\cite{MaeKut79}.
In terms of the the valence contribution, $N_e$, the principal resonance line oscillator
strength $f_{3s,3p}$, and the excitation energies of the first and second resonance transitions, 
respectively, $E_{3s,3p}$ and $E_{3s,4p}$,
they give
\begin{equation}
\alpha(0) = \frac{f_{3s,3p}}{E^2_{3s,3p}} + \frac{N_e-f_{3s,3p}}{2 E^2_{3s,4p}}
\end{equation}
and an uncertainty estimate for the polarizability 
\begin{equation}
\Delta \alpha(0) =\frac{N_e-f_{3s,3p}}{2 E^2_{3s,4p}}.
\end{equation}
Using the present  adopted value $f_{3s,3p}=1.75$, calculated
value  $N_e=2.16$, and transition
energies~\cite{KelPod08} $E_{3s,3p}= 0.159\,705$ and $E_{3s,4p}=0.224\,840$, yields
an estimate $\alpha(0)=72.7 \pm 4$.
Likewise, using their formula for estimating the uncertainty of $f_{3s,3p}$, given
the present calculated value
$\alpha(0)=72.2$, yields $f_{3s,3p}= 1.79 \pm 0.07$.
The formulae of Reshetnikov~\textit{et al.}~\cite{ResCurBro08} demonstrate 
that the present results are mutually consistent, but the estimates obtained are not sufficiently
precise to allow selection of a particular value of $f_{3s,3p}$
from the many available values, see Appendix.

The availability of a more accurate measurement
of $\alpha(0)$ and a definitive measurement of the principal resonance line
lifetime would significantly improve the present model~\cite{KhaBabDal97}.
Nevertheless, the values adopted here, in particular
$f_{3s,3p}=1.75$, generate sum rules that are consistent and not
in contradiction with other major studies,
while a value of $f_{3s,3p} \geq 1.8$ is inconsistent.

\section{Dynamic electric dipole polarizability function}\label{alpha-omega}

The dynamic electric dipole polarizability 
function at imaginary frequencies is constructed using the discrete and continuum
oscillator strength data as assembled in Sec.~\ref{OSD}.

The continuum oscillator strength distribution is given in terms of the photoionization
cross section $\sigma(E)$ by
\begin{equation}
\frac{df}{dE} = \frac{\sigma(E)}{2\pi^2 \alpha_{\mathrm{fs}}}, \quad E > 0.281 ,
\end{equation}
with $\alpha_{\mathrm{fs}}$ the fine structure constant, 
and the dynamic dipole polarizability at imaginary energy is
\begin{equation}
\label{eq-alpha-omega}
\alpha(i\omega) = \sum_n \frac{f_n}{(E_n- E_0)^2 + \omega^2}+ \int dE\,\frac{df/dE}{E^2 + \omega^2} .
\end{equation}

\begin{figure}
\includegraphics[scale=.75]{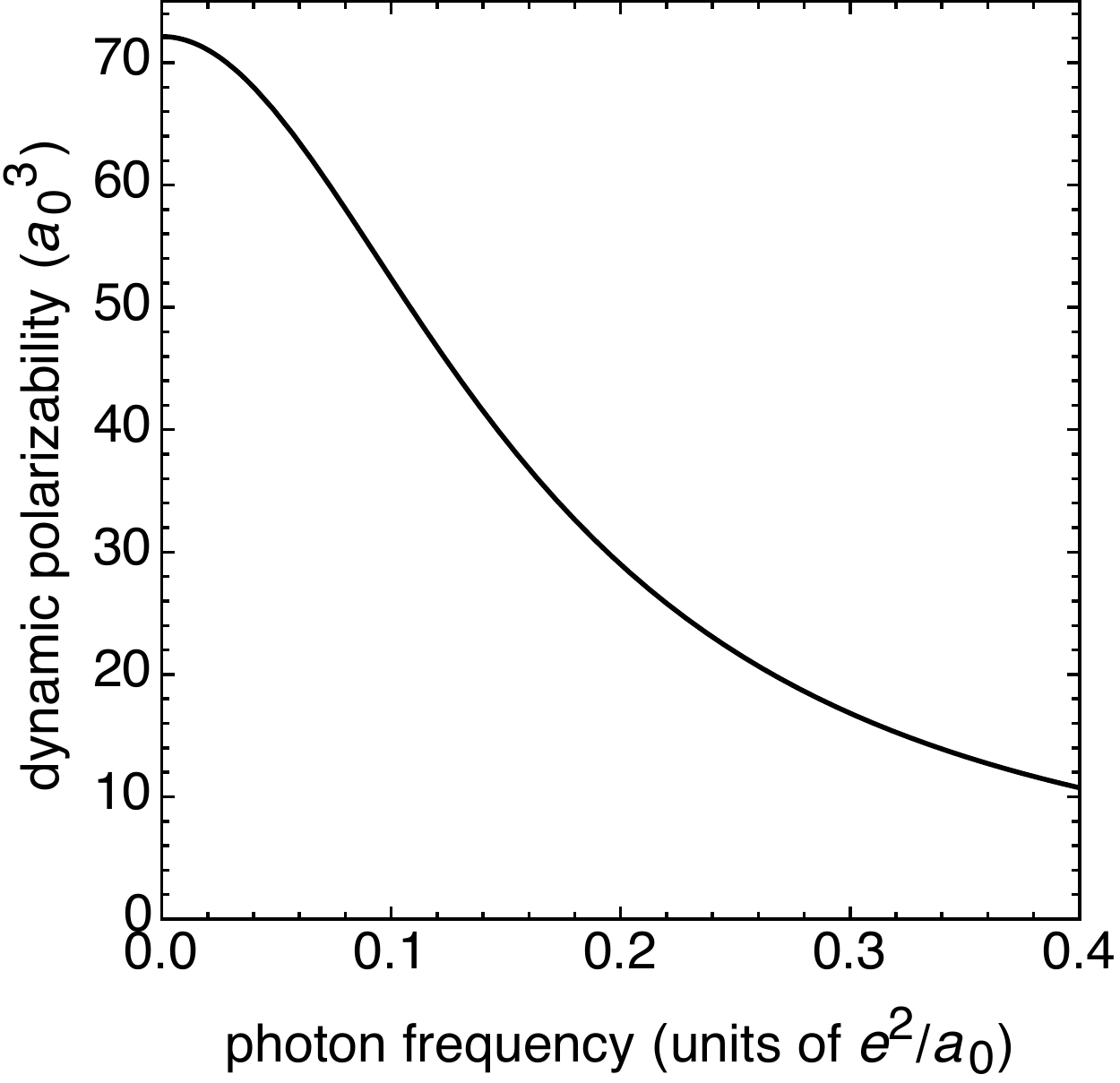}
\caption{The dynamic
dipole polarizability function $\alpha(i\omega)$
from the present calculations. 
\label{fig-pol-low}}
\end{figure}

\begin{figure}
\includegraphics[scale=.75]{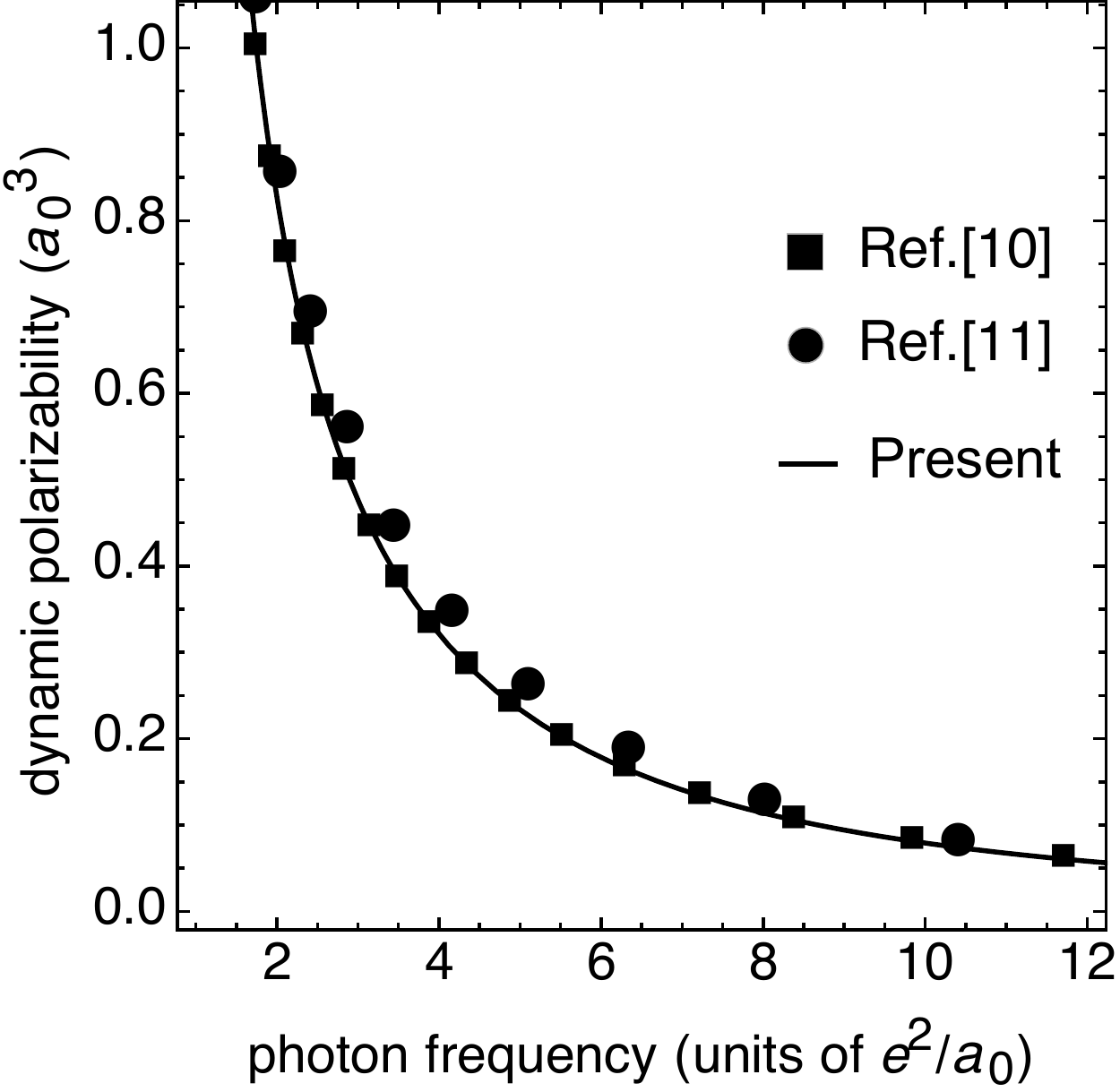}
\caption{For the dynamic
dipole polarizability function $\alpha(i\omega)$,  comparison  between the present values (line), the 
configuration interaction with semi-empirical core-valence interaction
(CICP) values
from Ref.~\cite{JiaMitChe15} (circles), and the configuration interaction
and many-body perturbation theory with core interactions (CI-MBPT) values
from Ref.~\cite{DerPorBab10} (squares). 
\label{fig-allthree}}
\end{figure}

\begin{figure}
\includegraphics[scale=.75]{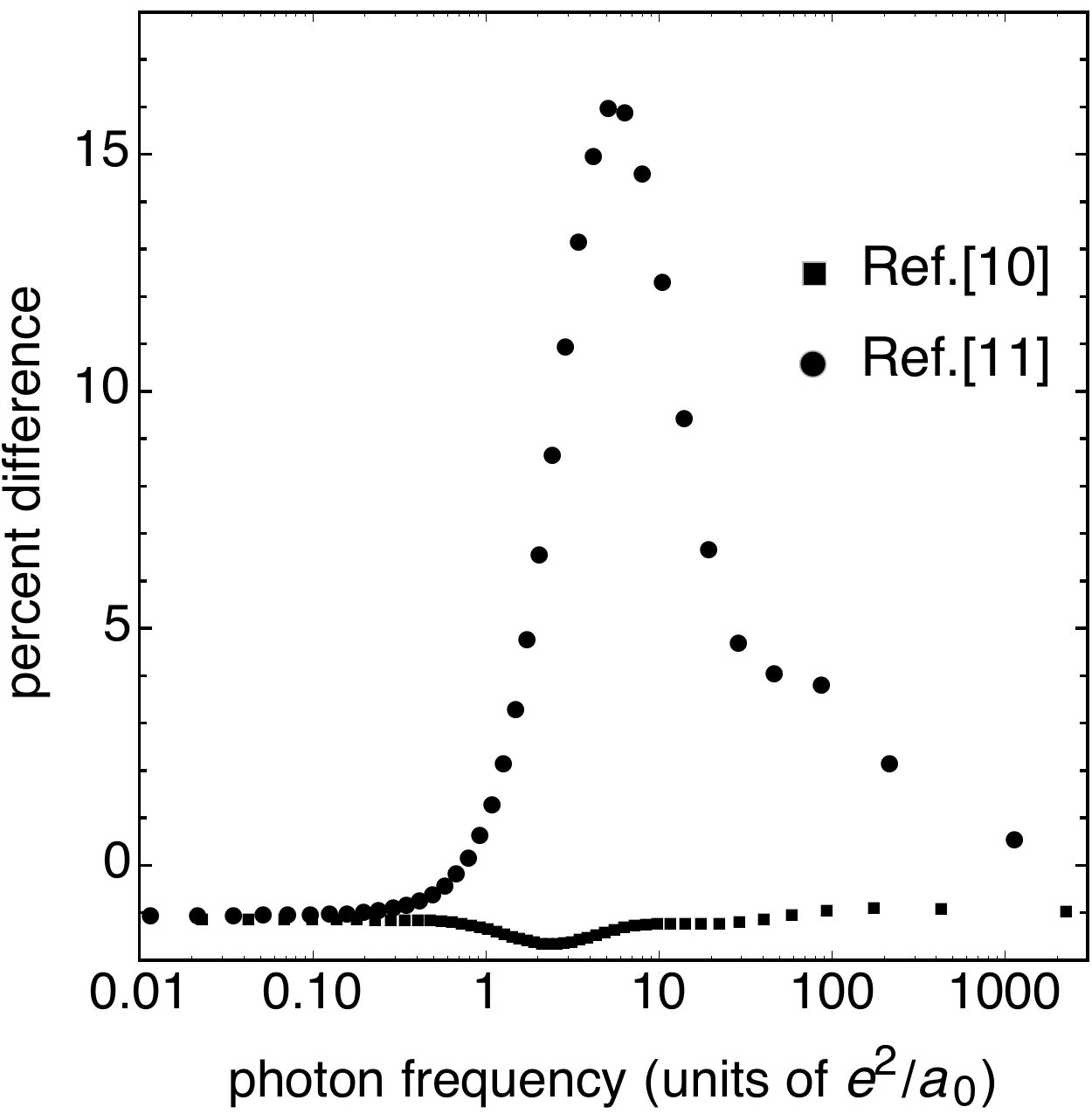}
\caption{For the dynamic
dipole polarizability function $\alpha(i\omega)$, percent difference comparison  between the present values and the 
CI-MBPT values
from Ref.~\cite{DerPorBab10} and between the
CICP values
from Ref.~\cite{JiaMitChe15} . The quantity plotted
is $100\times [( \textrm{other})- (\textrm{present} )]/\textrm{present}$,
where ``other'' is either Ref.~\cite{DerPorBab10} or~\cite{JiaMitChe15}. The filled squares are
the percent differences of the present values from the CI-MBPT values 
and the filled circles are the percent differences of
the present values from the CICP values.
The plotted values show that the present $\alpha(i\omega)$ is within several 
percent of the CI-MBPT values across the range of energies 
and show that the CICP values are larger 
for photon energies between roughly 1 and 200 atomic units.
\label{fig-alpha-diffs}}
\end{figure}

The  function  $\alpha (i\omega)$ resulting
from the analysis in Sec.~\ref{OSD} is shown
at low energies in Fig.~\ref{fig-pol-low}.
It may be compared with the calculations of Derevianko~\textit{et al.}~\cite{DerPorBab10}
and those of  Jiang~\textit{et al}~\cite{JiaMitChe15}.
The present function  $\alpha (i\omega)$ was evaluated at
the fifty energies  corresponding to
the energies $\omega_k$  of a 50-point quadrature, as listed in Table~A of Ref.~\cite{DerPorBab10},
and the  energies for a 40-point quadrature 
listed in Table~C of Jiang~\textit{et al.}
In Fig.~\ref{fig-allthree} the data are plotted.
Agreement is very good between the present results
and the CI-MBPT results of Ref.~\cite{DerPorBab10}.
There are noticeable discrepancies between the present results
and the CICP results of Ref.~\cite{JiaMitChe15}.
To further investigate the discrepancies, 
in Fig.~\ref{fig-alpha-diffs} the percentage difference between the values
from 
the functions given in Ref.~\cite{DerPorBab10} or Ref.~\cite{JiaMitChe15} 
and the present values
are shown.
The present model and the calculations
of Ref.~\cite{DerPorBab10} agree within
several percent at all energies.
The present values are larger than those of Ref.~\cite{DerPorBab10} at low energies,
due to the larger principal oscillator strength adopted herein. 
The several percent agreement across all energies is satisfactory
and a more detailed analysis might await an accurate
experimental value for $\alpha(0)$ and a definitive measurement of
the principal resonance line oscillator strength.
In contrast, 
the differences between the present model
and the model of Ref.~\cite{JiaMitChe15} 
are made apparent in Fig.~\ref{fig-alpha-diffs} showing that the CICP model
of Ref.~\cite{JiaMitChe15} yields larger values for $\alpha(i\omega)$ 
in the energy range of 1 to 200 au.
The difference may arise due to the choice of ``effective'' core oscillator strengths in the CICP model~\cite{MitBro03,JiaMitChe15}.
The percent difference peaks at about 16~\% around 5--6 $e^2/a_0$ (135--160~eV) 
placing the missing oscillator strength of the CICP model in the inner $s$ shells,
where the ``effective'' oscillator strengths are placed to model
inner shell absorption~\cite{MitBro03,JiaMitChe15}.

\section{Applications}

\subsection{van der Waals coefficient $C_6$}\label{C6}

The long-range potential energy between two Mg atoms  separated by
a distance $R$ is $-C_6/R^6$, where $C_6$ is given by Eq.~(\ref{vdW-formula}).
For the van der Waals coefficient, I find $C_6 = 642.4$ by evaluating
Eq.~(\ref{vdW-formula}) using the quadrature method of Ref.~\cite{DerPorBab10}, 
\begin{equation}
C_6 \approx \frac{3}{\pi} \sum_{j=1}^{50} w_j \alpha^2(i\omega_j),
\end{equation}
using the present values of $\alpha(i\omega)$ evaluated
at the energies $\omega_j$ and with the weights $w_j$ given in Table~A of Ref.~\cite{DerPorBab10}.
Porsev and Derevianko~\cite{PorDer02} 
quote accuracy of 2\% or better for their value $C_6=627(12)$ obtained 
using a semi-empirical hybrid relativistic many-body perturbation theory (CI-MBPT) approach. 
The present value lies just above their range, mainly corresponding
to their principal resonance line oscillator strength of 1.73, compared
to the present adopted value of 1.75, thus,  
$642 \times (1.73/1.75)^2 \approx 627$.
The present result  improves  upon the earlier empirical estimate  of $683(35)$ by Stwalley~\cite{Stw71}.
A more detailed survey and comparison of other determinations of $C_6$ is given in the Appendix.

\subsection{Atom-wall coefficient $C_3$}\label{C3}

The long-range potential energy of an Mg atom at distance $z$ from a perfectly conducting
wall is $-C_3/z^3$, where $C_3$ is given by Eq.~(\ref{atom-wall-formula}).
Mitroy and Bromley~\cite{MitBro03} calculated  $C_3 = 1.704$ using the 
CICP approach, while the
CI-MBPT value is $1.666$~\cite{DerPorBab10}.
Lonij \textit{et al.}~\cite{LonKlaHol11} gave an approximate value of $1.51$ using a limited 4-parameter model
for the dynamic polarizability. 
The value of $C_3$ is known to be sensitive to the completeness
of the description of the core electrons~\cite{FroFisJon98,DerJohSaf99,LonKlaHol10}.

The present value is $C_3=1.69$ using $C_3={\textstyle{\frac{1}{8}}} S(-1)$ and the value of $S(-1)$ from 
Table~\ref{osc} and 1.687 using Eq.~(\ref{atom-wall-formula})
and the quadrature from Ref.~\cite{DerPorBab10},
\begin{equation}
C_3 \approx \frac{1}{4\pi} \sum_{j=1}^{50} w_j  \alpha(i \omega_j) .
\end{equation}
Both of the present values (sum rule and quadrature) are in larger than that of Ref.~\cite{DerPorBab10}.
The slightly larger value of $C_3$ from the CICP calculations,
Ref.~\cite{JiaMitChe15}, is consistent with the relatively larger values of $\alpha(i\omega)$,
as discussed in Sec.~\ref{alpha-omega} and shown in Figs.~\ref{fig-allthree} and \ref{fig-alpha-diffs}.

\subsection{Other properties}
The Axilrod-Teller-Muto coefficient $C_9$, Eq.~(\ref{Axilrod-Teller-formula}), 
characterizes the mutual long-range interaction potential of three atoms.
The value obtained by Mitroy and Bromley~\cite{MitBro03} is $33\,380$ and
that obtained by Porsev~\textit{et al.}~\cite{DerPorBab10} is $33\,241$. 
Using the dynamic polarizability function I evaluated,
Eq.~(\ref{Axilrod-Teller-formula}), using the quadrature of Ref.~\cite{DerPorBab10},
\begin{equation}
\label{C9-approx}
C_9 \approx \frac{3}{\pi} \sum_{j=1}^{50} w_j \alpha^3 (i\omega_j),
\end{equation}
and obtained $C_9=34\,480$.

The  larger value for $C_9$ found here mainly reflects the larger principal oscillator
strength 1.75 adopted compared
to the principal oscillator strengths
found in  Refs.~\cite{MitBro03} and \cite{DerPorBab10}.
The  oscillator strength appears as a cubic power in Eq.~(\ref{C9-approx})
through $\alpha(i\omega)$, see Eq.~(\ref{eq-alpha-omega}). 
For example, comparing to Ref.~\cite{DerPorBab10}, which used a principal
oscillator strength of 1.73,
scaling the present value I obtain 
$34\,450 \times (1.73/1.75)^3 = 33\,310$, which is within $0.2$~\%  of the value of Ref.~\cite{DerPorBab10}.

\section{Conclusion}
Experimental and theoretical data were assembled
and used to formulate the dynamic polarizability function for Mg.
I find that consistency in the sum rules can be achieved using   
the adopted value of the principal resonance line oscillator
strength to be $1.75$; lower than
the curated values of 1.83~\cite{Mor03} and 1.8~\cite{KelPod08},
but in agreement with  theoretical calculations.
Comparisons of the dynamic dipole polarizability functions
from the present work and those calculated using the CI-MBPT approach and the
CICP approach were presented. Good agreement (within several percent)
was found with the CI-MBPT results over all photon energies providing an
independent confirmation of the CI-MPBT approach for Mg~\cite{DerPorBab10}.
For the CICP method the differences were more pronounced,
approaching 16\% 
at energies around 5--6 $e^2/a_0$,  or about 135--160~eV,
indicating that the ``effective"  oscillator strengths of Refs.~\cite{MitBro03}
and \cite{JiaMitChe15} may not completely  model 
oscillator strengths corresponding to the core electrons.
To improve the present model it would be valuable to have more accurate experimental measurements of the polarizability
and a definitive measurement of the principal oscillator strength. 
Where sufficient and reliable data is available, the present methodology can be applied to other atoms.
\textit{Note added in proof: A recent calculation [Y. Singh, B. K. Sahoo, and B. P. Das, Phys. Rev. A \textbf{88}, 062504 (2013)], using a relativistic coupled cluster
theory with all singly and doubly excited configurations,
found a polarizability value of 72.54(50).
I thank Dr. Singh for communicating this result.}

\begin{acknowledgments}
Discussions with  C. Ballance, T. G. Lee, T. Gorczyca, C. F. Fischer, S. Manson, and M. Bromley 
are gratefully acknowledged.
This work was supported in part by grants for ITAMP from the National Science Foundation to the Smithsonian Institution 
and to Harvard University. 
\end{acknowledgments}

\appendix*
\section{Values from the literature}

In this Appendix, values for Mg of  the principal resonance transition oscillator 
strength, of the static electric dipole polarizability, and of the van der Waals constant  
are collected from the literature. 
Some earlier collections include~Refs.~\cite{Mit75a,MenZei87a,MocSpi88b,RayMuk89} for the principal oscillator strength,
Refs.~\cite{ReiMey76,MaeKut79,RayMuk89,MitBro03,HamHib08,MitSafCla10} for the  polarizability,
and  Ref.~\cite{MitBro03} for the van der Waals constant.

\begin{table*}
\caption{\label{osc-table} A comparison 
of results for the principal resonance transition absorption oscillator strength for Mg.  
Abbreviations for methods are defined in the text.
Where a calculation was carried out in the length gauge (LG) and the  velocity gauge (VG),
the simple average is listed and the LG and VG
values are listed as a table footnote.
}
\begin{ruledtabular}
\begin{tabular}{llll}
\multicolumn{1}{l}{Method}  & \multicolumn{1}{l}{$f$} &
\multicolumn{1}{l}{Source} & \multicolumn{1}{l}{Ref. (Year)} \\
\hline
MCDF-CV		& 1.709 		& J\"onsson and Fischer 			& \cite{JonFis97} (1997)\\
MCHF			& 1.717 			& J\"onsson, Fischer, \& Godefroid 	& \cite{JonFisGod99} (1999)\\
MP			& 1.717 			& Victor and Laughlin  			& \cite{VicLau73} (1973)\\
MCHF+BP		& 1.719		&  Fischer, Tachiev, \& Irimia  	& \cite{FroFisTac06} (2006) \\
RCI+Breit		& 1.722		& Cheng \textit{et al.} 			& \cite{CheGaoQin11} (2011)  \\
CIDF-CP		& 1.72		& Stanek, Glowacki, \& Migdalek 	& \cite{StaGloMig96} (1996)\\
CI+MBPT\footnote{Mean of LG and VG line strength with experimental transition energy from \protect\cite{KelPod08}}
			& 1.724			& Savukov and Johnson 			& \cite{SavJoh02} (2002) \\
CI\footnote{Mean of LG 1.72 and VG 1.73}			
 			& 1.725			& Mengali and Moccia 			& \cite{MenMoc96b} (1996) \\
CI\footnote{Mean of LG 1.746 and VG 1.717}		
			& 1.73			& Nesbet and Jones 				& \cite{NesJon77} (1977)\\
CI+MBPT\footnote{Line strength with experimental transition energy from \protect\cite{KelPod08}}
			& 1.73(2)     		& Derevianko and Porsev 		& \cite{DerPor11} (2011)\\
CICP			& 1.732		& Mitroy and Bromley 			& \cite{MitBro03} (2003) \\
CI 			& 1.735			& Hamonou and Hibbert 			& \cite{HamHib08} (2008) \\
CI\footnote{Mean of LG 1.773 and VG 1.701}			
			& 1.737			& Weiss 						& \cite{Wei67} (1967) \\
MCHF-CV		& 1.738		& Zatsarinny \textit{et al.} 		& \cite{ZatBarGed09} (2009)\\
MCHF\footnote{Mean of LG 1.757 and VG 1.736}
    			& 1.747			& Fischer 						& \cite{Fis75} (1975) \\
CI-frozen core   & 1.75	        & Chang and Tang 				& \cite{ChaTan90} (1990)\\
CI-CV\footnote{Mean of LG 1.76 and VG 1.75}
			& 1.755	       	& Moccia and Spizzo 			& \cite{MocSpi88b} (1988)\\
CI-frozen core   &  1.76 		& Saraph 						& \cite{Sar76} (1976) \\
NIST adopted\footnote{Cited as Weiss, private communication. Using the CI value for the line strength from Weiss (1967) and the measured transition energy yields 1.77. }	& 1.8  	&  Kelleher and Podobedova  		& \cite{KelPod08} (2008) \\
Experiment\footnote{Weighted average of ten experimental values as of 2003.}
                                & 1.83(3)        & Morton                                           & \cite{Mor03} (2003) \\
\end{tabular}
\end{ruledtabular}
\end{table*}

For the oscillator strength, as discussed in the Sec.~\ref{OSD}, 
it was noted in 
several recent papers~\cite{JonFisGod99,ZatBarGed09} that 
in general the most sophisticated theoretical calculations lie several percent below
the  published experimental values, see also earlier similar comments
in Refs.~\cite{VicLau73}, \cite{VicSteLau76}, and \cite{MenZei87a}.
Also, it was noted that the ``best'' calculations lie below~\cite{ZatBarGed09} the adopted value of $1.8$ in the NIST tabulation~\cite{KelPod08}.
As shown in Table~\ref{osc-table},  the configuration interaction (CI),
multi-configuration Dirac-Fock (MCDF), and  multi-configuration Hartree-Fock with
Breit-Pauli interactions (MCHF+BP) calculations, are all in general agreement;
the latter two methods include relativistic effects.
In addition, the semi-empirical model potential (MP) calculation of Victor and Laughlin~\cite{VicLau73} agrees
with the MCHF calculation.
In the Table, where treatment of core-valence correlation
was included the suffix CV is appended.
In addition, the relativistic configuration interaction with the Breit interaction (RCI+Breit ) of Ref.~\cite{CheGaoQin11}
are in close agreement with the  configuration interaction Dirac-Fock with core polarization (CIDF-CP) calculation
of Ref.~\cite{StaGloMig96}.
CI+MBPT methods use complete relativistic CI calculations for the valence electrons in a frozen
core combined with MBPT to account for core-valence interactions.
The CI frozen core calculations of 
Saraph~\cite{Sar76} and Chang and Tang~\cite{ChaTan90} and the CI-CV calculations
of Moccia and Spizzo~\cite{MocSpi88b} are in close agreement 
with values between 1.75 and 1.76.
There is substantial theoretical evidence for a value of the principal oscillator strength around 1.75.

Many calculations of the static polarizability $\alpha(0)$  are available.
There are several good tables containing summaries of other earlier works.
In particular, from 1976, Reinsch and Meyer~\cite{ReiMey76} and from 1991,
Archibong and Thakkar~\cite{ArcTha91}, see also
Schwerdtfeger~\cite{Sch06}.
Thakkar and Lupinetti~\cite{ThaLup06} recommend a theoretical value of $71.22\pm 0.36$, which
includes a relativistic correction of $-0.35$, and Chu and Dalgarno~\cite{ChuDal04} recommend 71.

Table~\ref{alpha}
lists some of the results from the literature.
The quantum defect theory (QDT) value from Chernov \textit{et al.}~\cite{CheDorKre05}
models the response of the valence electrons only.
The pseudopotential (PP) \cite{MaeKut79} and model potential (MP) calculations ~\cite{VicSla74,Pat00} 
model the response of the 
valence electrons with inclusion of effective  potentials to treat the core electrons.
An effective core potential is used similarly  in the  configuration interaction core potential (CICP)
calculation of M\"{u}ller, Flesch, and Meyer~\cite{MulFleMey84}.
The CICP calculation of Mitroy and Bromley~\cite{MitBro03} utilizes a model
that treats core excitation using effective oscillator strengths
designed to reproduce the core polarizability.
The  multi-reference configuration interaction (MRCI) calculation of Partridge~\textit{et al.}~\cite{ParBauPet90}
gives 71.2. They also calculated a CI-CV value of  70.3 (not listed in Table~\ref{alpha}), which is in good agreement
with the similar calculation of 70.9 by Hamanou and Hibbert~\cite{HamHib08},
but the average value of 74.37 from the CI-CV calculations by Moccia and Spizzo~\cite{MocSpi88c} is
significantly larger.
The coupled cluster double-excitation with contributions
of single and triple excitations [CCD+ST (CCD)] model of Castro and Canuto~\cite{CasCan93} yields
a value of 70.89, somewhat lower than the MRCI value
and the fourth order many body perturbation theory  [MBPT(4)] value
of 71.7 calculated by Archibong and Thakkar~\cite{ArcTha91},
while the coupled cluster with single and double excitation-effective Hamiltonian (CCSD-EH)
approach of Stanton~\cite{Sta94} yields 72.2 using basis sets from Ref.~\cite{WidJoaPer91}, denoted WMR in the Table.
The  pseudo-natural orbital coupled electron pair approximation (PNO-CEPA)
calculation of Reinsch and Meyer~\cite{ReiMey76} 
is close to the CI+MBPT calculation of Porsev and Derevianko~\cite{PorDer06}.
Two  time-dependent density functional theory (TDDFT) calculations are included in Table~\ref{alpha}.
Using time-dependent DFT with
a self-interaction correction (TDDFT-SIC), Chu and Dalgarno~\cite{ChuDal04} obtained 71.8 
and using the symmetry
adapted perturbation theory codes, SAPT(DFT), Patkowski~\textit{et al.}~\cite{PatPodSza07} obtained  73.27 for the polarizability.
An extensive table of values for the polarizability of Mg
resulting from various density functional theory functionals is given in Ref.~\cite{BasHesSal08}.
The R-matrix calculation of Robb~\cite{Rob75} is 75(5);
the relatively large value results because core-valence correlation
effects were not included~\cite{MaeKut79}.
Excluding 
the relatively large value from Ref.~\cite{MocSpi88c}, the CI, MRCI, CICP, and MBPT calculations fall in the range from 70.74 to 71.7.
Reshetnikov \textit{et al.}~\cite{ResCurBro08} use a semi-empirical method
that utilizes a sum rule with constraints and error bars determined
using measured the lifetime and excitation energies.  Their value is 
$74.4(2.7)$, with the accuracy limited by the available input data.
The recent experiment of Ma~\textit{et al.} \cite{MaIndZha15} using a cryogenic molecular beam found a value
of $59(15)$, which is not yet sufficiently accurate to test the calculations against.

\begin{table*}
\caption{\label{alpha} The static electric dipole polarizability (in units of $a_0^3$) for Mg in the ground state. 
Abbreviations for methods are defined in the text.
}
\begin{ruledtabular}
\begin{tabular}{llll}
\multicolumn{1}{l}{Method}  & \multicolumn{1}{l}{$\alpha(0)$} &
\multicolumn{1}{l}{Source} & \multicolumn{1}{l}{Ref. (Year)} \\
\hline
Experiment (cryogenic molecular beam)
                           		& 59(15)     		& Ma \textit{et al.} 			& \cite{MaIndZha15} (2015) \\
QDT				& 69.54			& Chernov \textit{et al.}		& \cite{CheDorKre05} (2005) \\
PP		 			& 70.5   			&  Maeder and Kutzelnigg  	& \cite{MaeKut79} (1979) \\
CICP				& 70.74(71)		& M\"uller, Flesch, \& Meyer 	& \cite{MulFleMey84} (1984)\\
CCD+ST (CCD)		& 70.89 			& Castro and Canuto 		& \cite{CasCan93} (1993)\\
CI					& 70.90 			& Hamanou and Hibbert 		& \cite{HamHib08} (2008)\\
MRCI 				& 71.2  			& Partridge \textit{et al.} 		& \cite{ParBauPet90} (1990) \\
CICP				& 71.35			& Mitroy and Bromley 		& \cite{MitBro03} (2003) \\
CI+MBPT	      		& 71.3(7)		& Porsev and Derevianko 	& \cite{PorDer06} (2006) \\
PNO-CEPA       		& 71.32	 		& Reinsch and Meyer 		& \cite{ReiMey76} (1976) \\ 
MBPT(4) 			& 71.70    		& Archibong and Thakkar 	& \cite{ArcTha91} (1991) \\
TDDFT-SIC			& 71.8 			& Chu and Dalgarno 			& \cite{ChuDal04} (2004) \\
MP	       				& 72.0  			& Patil  						& \cite{Pat00} (2000)  \\
MP					& 72.1			& Victor and Slavsky			& \cite{VicSla74} (1974)\\
CCSD-EH (WMR)		& 72.2			& Stanton 					&  \cite{Sta94} (1994)\\
SAPT (DFT)			& 73.27			& Patkowski \textit{et al.} 		& \cite{PatPodSza07} (2007)\\
CI-CV\footnote{Mean of  LG 74.7 and VG 74.03} 
					& 74.37    		& Moccia and Spizzo 			& \cite{MocSpi88c} (1988)\\
Sum rule				& 74.4(2.7)       	& Reshetnikov \textit{et al.}   	& \cite{ResCurBro08} (2008) \\
R-matrix\footnote{Does not include core valence according to Ref.~\protect\cite{MaeKut79}}   
					& 75(5)    	& Robb 						&\cite{Rob75}  (1975)\\
\end{tabular}
\end{ruledtabular}
\end{table*}

Table~\ref{vdW} presents a collection of van der Waals constant values from the literature
and a significant range is apparent,
though the CI+MBPT and CICP calculations, which include models of 
core electron excitations,  are in good agreement.
The value $620(5)$ from the model potential (MP) calculation
of Santra, Christ, and Greene~\cite{SanChrGre04}  is close 
close to the  pseudopotential (PP) calculation of 618.4
from Maeder and Kutzelnigg~\cite{MaeKut79}, both of which 
include  effective potentials to account for the presence of core electrons,
but don't fully include their excitations.
The MP calculation from Patil~\cite{Pat00}, however, is significantly larger, at 648.
Three DFT calculations are listed in Table~\ref{vdW}.
Hult et al.~\cite{HulRydLun99} introduced a local dielectric function
and cutoff on the interaction volume and obtained 615.
In contrast, Chu and Dalgarno used time-dependent DFT with a self-interaction
correction (TDDFT-SIC) and an empirical correction to obtain 626 with an estimated
uncertainty of 1\%.
Patkoswki~\textit{et al.}~\cite{PatPodSza07} used the symmetry adapted
perturbation theory codes, SAPT(DFT), and obtained 635.
The calculation of $C_6$ by Stanton~\cite{Sta94} used a quadrature and values of the dynamic polarizability
at imaginary frequencies calculated using the CCSD-EH coupled cluster approach with
the basis sets from~\cite{WidJoaPer91}.
The CI-CV calculations of Moccia and Spizzo~\cite{MocSpi88c} in the 
velocity gauge (VG) and in the length gauge (LG)  are
substantially larger than the other listed calculations.
Robb~\cite{Rob75}  estimated his R-matrix calculation to be accurate to 10~\%.
Stwalley~\cite{Stw71} used an empirically constructed polarizability function
to calculate $C_6$. The large  value for $C_6$ corresponds to the choice of 1.82 for the principal
oscillator strength.
In Ref.~\cite{KnoRuhTie13} it was found that a 
2\% uncertainty in $C_6$ leads to an uncertainty of no more than 0.3~nm in
the scattering length for ${}^{24}\textrm{Mg}_2$.

\begin{table*}
\caption{\label{vdW} The dispersion constant (in units of $e^2 a_0^5$) for the Mg dimer 
from various references. Abbreviations for methods are defined in the text.
}
\begin{ruledtabular}
\begin{tabular}{llll}
\multicolumn{1}{l}{Method}  & \multicolumn{1}{l}{$C_6$} &
\multicolumn{1}{l}{Source} & \multicolumn{1}{c}{Ref. (Year)} \\
\hline
DFT					& 615 		& Hult \textit{at al.} 		& \cite{HulRydLun99} (1999)\\
PP				       & 618.4  		&  Maeder and Kutzelnigg	&  \cite{MaeKut79} (1979)\\
MP\footnote{Does not include core contributions}
                		       & 620(5) 	&Santra, Christ, \& Greene	& \cite{SanChrGre04} (2004)\\
TDDFT-SIC\footnote{Listed in Table VII as ``corrected,'' corresponding to an empirical rescaling.}
		    			& 626(6) 	& Chu and Dalgarno		& \cite{ChuDal04}  (2004)\\
CI+MBPT			& 627(12)	& Porsev and Derevianko 	& \cite{PorDer06} (2006)\\
CICP				& 629.5 		& Mitroy and Bromley		& \cite{MitBro03} (2003)\\
MP					& 632.27		& Victor and Slavsky		& \cite{VicSla74} (1974)\\
SAPT(DFT)			& 635		& Patkowski \textit{et al.}	& \cite{PatPodSza07} (2007)\\
CCSD-EH (WMR)	       & 648 		& Stanton 				&  \cite{Sta94} (1994)\\
MP					& 648		& Patil 				& \cite{Pat00} (2000) \\
CI-CV (VG)			& 658.1 		& Moccia and Spizzo  		& \cite{MocSpi88c} (1988)\\
CI-CV (LG)			& 670.9 		& Moccia and Spizzo  		& \cite{MocSpi88c} (1988)\\
Empirical   			& 683(35)	& Stwalley 			& \cite{Stw71} (1971)\\
R-matrix         		& 689(70) 		& Robb 				& \cite{Rob75} (1975)
\end{tabular}
\end{ruledtabular}
\end{table*}


\begin{thebibliography}{100}

\bibitem{TieKotJul02}
E. Tiesinga, S. Kotochigova, and P.~S. Julienne, Phys. Rev. A {\bf 65},  042722
   (2002).

\bibitem{KnoRuhTie13}
H. Kn\"ockel, S. R\"uhmann, and E. Tiemann, J. Chem. Phys. {\bf 138},  094303
  (2013).

\bibitem{KnoRuhTie14}
H. Kn\"ockel, S. R\"uhmann, and E. Tiemann, Eur. Phys. J. D {\bf 68},  293
  (2014).

\bibitem{SolZucHut09}
P. Soldan, P.~S. Zuchowski, and J.~M. Hutson, Discuss. Faraday Soc. {\bf 142},
  191  (2009).

\bibitem{GonMarMay11}
M.~L. Gonz\'alez-Mart\'inez and J.~M. Hutson, {Phys. Rev. A} {\bf {84}},
  052706  ({2011}).

\bibitem{LonKlaHol11}
V.~P.~A. Lonij, C.~E. Klauss, W.~F. Holmgren, and A.~D. Cronin, J. Phys. Chem.
  A {\bf 115},  7134  (2011).

\bibitem{VidMadHui13}
A. Vidal-Madjar, C.~M. Huitson, V. Bourrier, J.-M. D\'esert, G. Ballester, A.
  Lecavelier~des Etangs, D.~K. Sing, D. Ehrenreich, R. Ferlet, G. H\'ebrard,
  and J.~C. McConnell, Astron. Astroph. {\bf 560},  A54  (2013).

\bibitem{BouLecVid14}
V. Bourrier, A. Lecavelier~des Etangs, and A. Vidal-Madjar, Astron. Astroph.
  {\bf 565},  A105  (2014).

\bibitem{BouLecVid15}
V. Bourrier, A. Lecavelier~des Etangs, and A. Vidal-Madjar, Astron. Astroph.
  {\bf 573},  A11  (2015).

\bibitem{DerPorBab10}
A. Derevianko, S. Porsev, and J. Babb, At. Data Nucl. Data Tables {\bf 96},
  323  (2010).

\bibitem{JiaMitChe15}
J. Jiang, J. Mitroy, Y. Cheng, and M. Bromley, At. Data Nucl. Data Tables {\bf
  101},  158  (2015).

\bibitem{Amu90}
M.~Y. Amusia, {\em Atomic Photoeffect} (Plenum, New York, 1990), trans. K. T.
  Taylor.

\bibitem{AmuCheYar12}
M. {Amusia}, L. {Chernysheva}, and V. {Yarzhemsky}, {\em Handbook of
  Theoretical Atomic Physics Data for Photon Absorption, Electron Scattering,
  and Vacancies Decay} (Springer, Berlin, 2012).

\bibitem{Ber02}
J. Berkowitz, {\em Atomic and Molecular Photoabsorption: {Absolute} Total Cross
  Sections} (Academic Press, San Diego, 2002).

\bibitem{MasSta00}
M. Masili and A.~F. Starace, {Phys. Rev. A} {\bf {62}},  {033403}  ({2000}).

\bibitem{MidFalLis12}
T. Middelmann, S. Falke, C. Lisdat, and U. Sterr, Phys. Rev. Lett. {\bf 109},
  263004  (2012).

\bibitem{SheLemHin12}
J.~A. Sherman, N.~D. Lemke, N. Hinkley, M. Pizzocaro, R.~W. Fox, A.~D. Ludlow,
  and C.~W. Oates, Phys. Rev. Lett. {\bf 108},  153002  (2012).

\bibitem{BelSheLem12}
K. Beloy, J.~A. Sherman, N.~D. Lemke, N. Hinkley, C.~W. Oates, and A.~D.
  Ludlow, Phys. Rev. A {\bf 86},  051404  (2012).

\bibitem{StaBudFre06}
J.~E. Stalnaker, D. Budker, S.~J. Freedman, J.~S. Guzman, S.~M. Rochester, and
  V.~V. Yashchuk, Phys. Rev. A {\bf {73}},  {043416}  ({2006}).

\bibitem{BarStaLem08}
Z.~W. Barber, J.~E. Stalnaker, N.~D. Lemke, N. Poli, C.~W. Oates, T.~M.
  Fortier, S.~A. Diddams, L. Hollberg, C.~W. Hoyt, A.~V. Taichenachev, and
  V.~I. Yudin, Phys. Rev. Lett. {\bf 100},  103002  (2008).

\bibitem{Khr91}
I.~B. Khriplovich, {\em Parity Non-conservation in Atomic Phenomena} (Gordon
  and Breach, New York, 1991).

\bibitem{BasHesSal08}
R. Bast, A. Hesselmann, P. Sa\l{}ek, T. Helgaker, and T. Saue, {ChemPhysChem}
  {\bf 9},  445  (2008).

\bibitem{TkaSch09}
A. Tkatchenko and M. Scheffler, Phys. Rev. Lett. {\bf 102},  073005  (2009).

\bibitem{TkaDiSCar12}
A. Tkatchenko, R.~A. DiStasio, R. Car, and M. Scheffler, Phys. Rev. Lett. {\bf
  108},  236402  (2012).

\bibitem{TaoPerRuz13}
J. Tao, J.~P. Perdew, and A. Ruzsinszky, Int. J. Mod. Phys. B {\bf 27},
  1330011  (2013).

\bibitem{TouRebGou13}
J. Toulouse, E. Rebolini, T. Gould, J.~F. Dobson, P. Seal, and J.~G. Angyan, J.
  Chem. Phys. {\bf 138},  194106  (2013).

\bibitem{GoeHoh95}
D. Goebel and U. Hohm, Phys. Rev. A {\bf 52},  3691  (1995).

\bibitem{SarBeiShe06}
G.~S. Sarkisov, I.~L. Beigman, V.~P. Shevelko, and K.~W. Struve, Phys. Rev. A
  {\bf 73},  042501  (2006).

\bibitem{MaIndZha15}
L. Ma, J. Indergaard, B. Zhang, I. Larkin, R. Moro, and W.~A. de~Heer, Phys.
  Rev. A {\bf 91},  010501  (2015).

\bibitem{Sch06}
P. Schwerdtfeger,  in {\em Atoms, Molecules, and Clusters in Electric Fields},
  edited by G. Maroulis (Imperial College Press, London, 2006), Chap.~1, p.\ 1.

\bibitem{ThaLup06}
A.~J. Thakkar and C. Lupinetti,  in {\em Atoms, Molecules, and Clusters in
  Electric Fields}, edited by G. Maroulis (Imperial College Press, London,
  2006), Chap.~14, pp.\ 505--530.

\bibitem{MitSafCla10}
J. Mitroy, M.~S. Safronova, and C.~W. Clark, J. Phys. B {\bf 43},  202001
  (2010).

\bibitem{DimGer03}
S. Dimopoulos and A.~A. Geraci, Phys. Rev. D {\bf 68},  124021  (2003).

\bibitem{HarObrMcG05}
D.~M. Harber, J.~M. Obrecht, J.~M. McGuirk, and E.~A. Cornell, Phys. Rev. A
  {\bf 72},  033610  (2005).

\bibitem{MurHau09}
B. Murphy and L.~V. Hau, Phys. Rev. Lett. {\bf 102},  033003  (2009).

\bibitem{DerObrDzu09}
A. Derevianko, B. Obreshkov, and V.~A. Dzuba, Phys. Rev. Lett. {\bf 103},
  133201  (2009).

\bibitem{DzuDer10}
V.~A. Dzuba and A. Derevianko, J. Phys. B: At. Mol. Opt. Phys. {\bf 43},
  074011  (2010).

\bibitem{AroKauSah14}
B. Arora, H. Kaur, and B.~K. Sahoo, J. Phys. B {\bf 47},  155002  (2014).

\bibitem{JenLacDeK15}
U.~D. Jentschura, G. \L{}ach, M. De~Kieviet, and K. Pachucki, Phys. Rev. Lett.
  {\bf 114},  043001  (2015).

\bibitem{TaoPer14}
J. Tao and J.~P. Perdew, J. Chem. Phys. {\bf 141},  141101  (2014).

\bibitem{KhaBabDal97}
P. Kharchenko, J.~F. Babb, and A. Dalgarno, Phys. Rev. A {\bf 55},  3566
  (1997).

\bibitem{vanVer99}
F.~A. van Abeelen and B.~J. Verhaar, Phys. Rev. A {\bf 59},  578  (1999).

\bibitem{KnoSchSce11}
S. Knoop, T. Schuster, R. Scelle, A. Trautmann, J. Appmeier, M.~K. Oberthaler,
  E. Tiesinga, and E. Tiemann, Phys. Rev. A {\bf 83},  042704  (2011).

\bibitem{DerJohSaf99}
A. {Derevianko}, W.~R. {Johnson}, M.~S. {Safronova}, and J.~F. {Babb}, Phys.
  Rev. Lett. {\bf 82},  3589  (1999).

\bibitem{Mor03}
D.~C. {Morton}, Astrophys. J., Suppl. Ser. {\bf 149},  205  (2003).

\bibitem{KelPod08}
D.~E. Kelleher and L.~I. Podobedova, J. Phys. Chem. Ref. Data {\bf 37},  267
  (2008).

\bibitem{PorDer06}
S.~G. Porsev and A. Derevianko, Zh. Eksp. Teor. Fiz. {\bf 129},  227  (2006),
  [JETP \textbf{102}, 195 (2006)].

\bibitem{Mit75a}
C.~J. Mitchell, J. Phys. B {\bf 8},  25  (1975).

\bibitem{MenZei87a}
C. {Mendoza} and C.~J. {Zeippen}, Astron. Astroph. {\bf 179},  339  (1987).

\bibitem{RayMuk89}
D. Ray and P.~K. Mukherjee, J. Phys. B {\bf 22},  2103  (1989).

\bibitem{JonFis97}
P. J\"onsson and C.~F. Fischer, J. Phys. B {\bf 30},  5861  (1997).

\bibitem{HamHib08}
L. Hamonou and A. Hibbert, J. Phys. B {\bf 41},  245004  (2008).

\bibitem{DerPor11}
A. Derevianko and S.~G. Porsev, {\em Adv. At. Molec. Opt. Phys.}, edited by E.
  Arimondo, P.~R. Berman, and C.~C. Lin (Academic Press, Amsterdam, 2011),
  Vol.~60, Chap.~9, pp.\ 415 -- 459.

\bibitem{Fis75}
C.~F. Fischer, Can. J. Phys. {\bf 53},  338  (1975).

\bibitem{VicSteLau76}
G.~A. {Victor}, R.~F. {Stewart}, and C. {Laughlin}, Ap. J., Suppl. {\bf 31},
  237  (1976).

\bibitem{JonFisGod99}
P. J\"onsson, C.~F. Fischer, and M.~R. Godefroid, J. Phys. B {\bf 32},  1233
  (1999).

\bibitem{ZatBarGed09}
O. Zatsarinny, K. Bartschat, S. Gedeon, V. Gedeon, V. Lazur, and E. Nagy, Phys.
  Rev. A {\bf 79},  052709  (2009).

\bibitem{PorKozRak01}
S.~G. Porsev, M.~G. Kozlov, Y.~G. Rakhlina, and A. Derevianko, Phys. Rev. A
  {\bf 64},  012508  (2001).

\bibitem{WieSmiMil69}
W.~L. Wiese, J.~R. Fuhr, and B.~M. Miles, {\em Atomic Transition Probabilities,
  Vol. II: Sodium through Calcium, NSRDS-NBS Vol. 22} (US GPO, Washington,
  D.C., 1969).

\bibitem{Lur64}
A. Lurio, Phys. Rev. {\bf 136},  A376  (1964).

\bibitem{SmiGal66}
W.~W. Smith and A. Gallagher, Phys. Rev. {\bf 145},  26  (1966).

\bibitem{Wei67}
A.~W. Weiss, J. Chem. Phys. {\bf 47},  3573  (1967).

\bibitem{VicLau73}
G. Victor and C. Laughlin, Nucl. Instrum. Methods {\bf 110},  189   (1973).

\bibitem{Sar76}
H.~E. Saraph, J. Phys. B {\bf 9},  2379  (1976).

\bibitem{ChaTan90}
T.~N. Chang and X. Tang, J. Quant. Spectrosc. Radiat. Transfer {\bf 43},  207
  (1990).

\bibitem{DitMar53}
R.~W. {Ditchburn} and G.~V. {Marr}, Proc. Phys. Soc., London, Sect. A {\bf 66},
   655  (1953).

\bibitem{BurSea60}
A. {Burgess} and M.~J. {Seaton}, Mon. Not. R. Astron. Soc. {\bf 120},  121
  (1960).

\bibitem{BatAlt73}
G.~N. {Bates} and P.~L. {Altick}, J. Phys. B {\bf 6},  653  (1973).

\bibitem{ParReeTom76}
W.~H. Parkinson, E.~M. Reeves, and F.~S. Tomkins, J. Phys. B {\bf 9},  157
  (1976).

\bibitem{DesMan83}
P.~C. Deshmukh and S.~T. Manson, Phys. Rev. A {\bf 28},  209  (1983).

\bibitem{PreBurGar84}
J.~M. Preses, C.~E. Burkhardt, W.~P. Garver, and J.~J. Leventhal, Phys. Rev. A
  {\bf 29},  985  (1984).

\bibitem{RadJoh85}
V. Radojevi\ifmmode~\acute{c}\else \'{c}\fi{} and W.~R. Johnson, Phys. Rev. A
  {\bf 31},  2991  (1985).

\bibitem{YehLin85}
J. Yeh and I. Lindau, At. Dat. Nucl. Dat. Tables {\bf 32},  1  (1985).

\bibitem{FisSah87}
C.~F. Fischer and H.~P. Saha, Can. J. Phys. {\bf 65},  772  (1987).

\bibitem{MenZei87b}
C. {Mendoza} and C.~J. {Zeippen}, Astron. Astroph. {\bf 179},  346  (1987).

\bibitem{MocSpi88b}
R. Moccia and P. Spizzo, J. Phys. B {\bf 21},  1133  (1988).

\bibitem{Alt89}
Z. Altun, Phys. Rev. A {\bf 40},  4968  (1989).

\bibitem{VerYakBan93}
D.~A. {Verner}, D.~G. {Yakovlev}, I.~M. {Band}, and M.~B. {Trzhaskovskaya}, At.
  Dat. Nucl. Dat. Tables {\bf 55},  233  (1993).

\bibitem{ChiHua94}
H.-C. Chi and K.-N. Huang, Phys. Rev. A {\bf 50},  392  (1994).

\bibitem{FunYih01}
H.~S. Fung and T.~S. Yih, Nucl. Phys. A {\bf 684},  696C  (2001).

\bibitem{KimTay00}
D.-S. Kim and S.~S. Tayal, J. Phys. B: At. Mol. Opt. Phys. {\bf 33},  3235
  (2000).

\bibitem{WehLukJur07}
R. Wehlitz, D. Luki\'{c}, and P.~N. Jurani\'{c}, J. Phys. B {\bf 40},  2385  (2007).

\bibitem{HauKamKos88}
A. Hausmann, B. K\"ammerling, H. Kossmann, and V. Schmidt, Phys. Rev. Lett.
  {\bf 61},  2669  (1988).

\bibitem{WanWanZho10}
G. Wang, J. Wan, and X. Zhou, J. Phys. B {\bf 43},  035001  (2010).

\bibitem{PinBalAbd13}
M.~S. Pindzola, C.~P. Ballance, S.~A. Abdel-Naby, F. Robicheaux, G.~S.~J.
  Armstrong, and J. Colgan, J. Phys. B {\bf 46},  035201  (2013).

\bibitem{LeeBalAbd15}
T.~G. Lee, C.~P. Ballance, S.~A. Abdel-Naby, J.~L. King, T.~W. Gorczyca, and
  M.~S. Pindzola, J. Phys. B {\bf 48},  065201  (2015).

\bibitem{HenLeeTan82}
B.~L. Henke, P. Lee, T.~J. Tanaka, R.~L. Shimabukuro, and B.~K. Fujikawa, At.
  Dat. Nucl. Dat. Tables {\bf 27},  1  (1982).

\bibitem{Stw71}
W.~C. Stwalley, J. Chem. Phys. {\bf 54},  4517  (1971).

\bibitem{ResCurBro08}
N. Reshetnikov, L.~J. Curtis, M.~S. Brown, and R.~E. Irving, Phys. Scr. {\bf
  77},  015301  (2008).

\bibitem{MaeKut79}
F. Maeder and W. Kutzelnigg, Chem. Phys. {\bf 42},  95  (1979).

\bibitem{KutMayTho02}
M. Kutzner, V. Maycock, J. Thorarinson, E. Pannwitz, and J.~A. Robertson, Phys.
  Rev. A {\bf 66},  042715  (2002).

\bibitem{HasAbdNab14}
M.~F. Haso{\u g}lu, S.~A. Abdel-Naby, E. Gatuzz, J. Garc{\'\i}a, T.~R. Kallman,
  C. Mendoza, and T.~W. Gorczyca, Astrophys. J., Suppl. Ser. {\bf 214},  8
  (2014).

\bibitem{BanSlaMat82}
M. Banna, A. Slaughter, R. Mathews, R. Key, and S. Ballina, Chem. Phys. Lett.
  {\bf 92},  122   (1982).

\bibitem{NasManDes89}
G. Nasreen, S.~T. Manson, and P.~C. Deshmukh, Phys. Rev. A {\bf 40},  6091
  (1989).

\bibitem{PalOvs04}
V.~G. Pal'chikov and V.~D. Ovsiannikov, Quantum Electronics {\bf 34},  412
  (2004).

\bibitem{OvsPalKat06}
V.~D. Ovsyannikov, V.~G. Pal'chikov, H. Katori, and M. Takamoto, Quantum
  Electron. {\bf 36},  3  (2006).

\bibitem{MitBro03}
J. Mitroy and M.~W.~J. Bromley, Phys. Rev. A {\bf 68},  052714  (2003).

\bibitem{PorDer02}
S.~G. Porsev and A. Derevianko, Phys. Rev. A {\bf 65},  020701  (2002).

\bibitem{FroFisJon98}
C. Froese~Fischer, P. J\"onsson, and M. Godefroid, Phys. Rev. A {\bf 57},  1753
   (1998).

\bibitem{LonKlaHol10}
V.~P.~A. Lonij, C.~E. Klauss, W.~F. Holmgren, and A.~D. Cronin, Phys. Rev.
  Lett. {\bf 105},  233202  (2010).

\bibitem{ReiMey76}
E.-A. Reinsch and W. Meyer, Phys. Rev. A {\bf 14},  915  (1976).

\bibitem{FroFisTac06}
C. Froese~Fischer, G. Tachiev, and A. Irimia, At. Dat. Nucl. Dat. Tables {\bf
  92},  607  (2006).

\bibitem{CheGaoQin11}
C. Cheng, X. Gao, B. Qing, X.-L. Zhang, and J.-M. Li, Chin. Phys. B {\bf 20},
  033103  (2011).

\bibitem{StaGloMig96}
M. Stanek, L. Glowacki, and J. Migdalek, J. Phys. B {\bf 29},  2985  (1996).

\bibitem{SavJoh02}
I.~M. Savukov and W.~R. Johnson, Phys. Rev. A {\bf 65},  042503  (2002).

\bibitem{MenMoc96b}
S. Mengali and R. Moccia, J. Phys. B {\bf 29},  1613  (1996).

\bibitem{NesJon77}
R.~K. Nesbet and H.~W. Jones, Phys. Rev. A {\bf 16},  1161  (1977).

\bibitem{ArcTha91}
E.~F. Archibong and A.~J. Thakkar, Phys.~Rev.~A {\bf 44},  5478  (1991).

\bibitem{ChuDal04}
X. Chu and A. Dalgarno, J. Chem. Phys. {\bf 121},  4083   (2004).

\bibitem{CheDorKre05}
V.~E. Chernov, D.~L. Dorofeev, I.~Y. Kretinin, and B.~A. Zon, Phys. Rev. A {\bf
  71},  022505  (2005).

\bibitem{VicSla74}
G.~A. Victor and D.~B. Slavsky, J. Chem. Phys. {\bf 61},  3484  (1974).

\bibitem{Pat00}
S. Patil, Eur. Phys. J. D {\bf 10},  341  (2000).

\bibitem{MulFleMey84}
W. M{\"u}ller, J. Flesch, and W. Meyer, J. Chem. Phys. {\bf 80},  3297  (1984).

\bibitem{ParBauPet90}
H. Partridge, C.~W. Bauschlicher~Jr., L.~G.~M. Pettersson, A.~D. McLean, B.
  Liu, M. Yoshimine, and A. Komornicki, J. Chem. Phys. {\bf 92},  5377  (1990).

\bibitem{MocSpi88c}
R. Moccia and P. Spizzo, J. Phys. B {\bf 21},  1145  (1988).

\bibitem{CasCan93}
M.~A. Castro and S. Canuto, Phys. Lett. A {\bf 176},  105  (1993).

\bibitem{Sta94}
J.~F. Stanton, Phys. Rev. A {\bf 49},  1698  (1994).

\bibitem{WidJoaPer91}
P. Widmark, B. Joakim~Persson, and B. Roos, Theo. Chim. ACTA {\bf 79},  419
  (1991).

\bibitem{PatPodSza07}
K. Patkowski, R. Podeszwa, and K. Szalewicz, J. Phys. Chem. A {\bf 111},  12822
   (2007).

\bibitem{Rob75}
W.~D. {Robb}, Chem. Phys. Lett. {\bf 34},  479  (1975).

\bibitem{SanChrGre04}
R. Santra, K.~V. Christ, and C.~H. Greene, Phys. Rev. A {\bf 69},  042510
  (2004).

\bibitem{HulRydLun99}
E. Hult, H. Rydberg, B.~I. Lundqvist, and D.~C. Langreth, Phys. Rev. B {\bf
  59},  4708  (1999).

\end{thebibliography}

\end{document}